# Material loss angles from direct measurements of broadband thermal noise

Maria Principe, Innocenzo M. Pinto, Vincenzo Pierro, Riccardo DeSalvo, and Ilaria Taurasi
*Waves Group, University of Sannio at Benevento, Benevento I-82100, Italy, INFN, LVC, and KAGRA*

Akira E. Villar, Eric D. Black, and Kenneth G. Libbrecht
*LIGO Laboratory, California Institute of Technology Mail Code 264-33,
Pasadena, California 91125, USA*

Christophe Michel, Nazario Morgado, and Laurent Pinard
*Laboratoire des Materiaux Avances, Universite Claude Bernard Lyon 1,
CNRS/IN2P3, 69622 Villeurbaune, Cedex, France*

**Abstract**

We estimate the loss angles of the materials currently used in the highly reflective test-mass coatings of interferometric detectors of gravitational waves, namely Silica, Tantala, and Ti-doped Tantala, from direct measurement of coating thermal noise in an optical interferometer testbench, the Caltech TNI. We also present a simple predictive theory for the material properties of amorphous glassy oxide mixtures, which gives results in good agreement with our measurements on Ti-doped Tantala. Alternative measurement methods and results are reviewed, and some critical issues are discussed.

## I. INTRODUCTION

A number of observatories based on optical interferometric detectors of gravitational waves (henceforth GW) have been already built (LIGO [1], GEO [2], Virgo [3], and TAMA [4]), are under construction (KAGRA, formerly LCGT [5]), or have been proposed (ACIGA [6] and INDIGO [7]). Second-generation upgrades of existing detectors have been implemented using new materials and technologies, to reduce their noise floor and improve their astrophysical reach [8].

Thermal noise in the high-reflectivity dielectric coatings of the test masses sets the limiting sensitivity of these instruments. Reducing coating thermal noise is essential if we want to reach the standard quantum-noise limit, and such a reduction is also a necessary prerequisite for any quantum nondemolition schemes to surpass this limit [9].

The coatings used in both first- and second-generation GW detectors consist of alternating layers of materials with a high and low index of refraction [9]. Coating materials presently in use belong to the class of amorphous glassy oxides [10] including, among others, $SiO_2$, $ZrO_2$, $HfO_2$, $TiO_2$, $Al_2O_3$, $Ta_2O_5$, and $Nb_2O_5$. Noise in these coatings originates from mechanical dissipation in the coating materials via a mechanism described by the fluctuation-dissipation theorem. On the basis of available evidence, dissipation in the bulk of the coating materials appears to be the dominant mechanism, and interfacial friction between coating layers and between coating and substrate is comparatively negligible [12].

Most estimates of material loss angles obtained so far are based on the measurement of the mechanical quality factor, or damping time of coated blades.

A physically sound and well-credited theory relates optical and mechanical properties of amorphous materials to the existence of asymmetric double-well potentials representing material defects. The complex frequency-dependent optical index and Young's modulus can be in principle obtained from the distributions of the potential barrier heights and height asymmetries [13], but so far, this theory has not yielded yet quantitative predictions of loss angles in the actual materials used in GW detectors. Until now, all attempts to synthesize coating materials with better optical and mechanical properties using glassy-oxide mixtures [14] have been essentially based on trial and error.

We present here the first extraction of the *individual* loss angles of the materials currently used in the mirror coatings of interferometric GW detectors, namely Silica, Tantala, and Titania-doped Tantala, based on the *direct* measurement of coating thermal noise in an interferometric (i.e., GW detector-like) setting (see Secs. II and III). A preliminary account of these findings was given in [15].

We also propose here for the first time in this field, to the best of our knowledge, a simple predictive model for the optical and mechanical properties of glassy-oxide mixtures based on effective-medium theory (EMT) (see Sec. IV). This model yields results in good agreement with our measurements on Titania doped Tantala-based coatings, discussed in Secs. III A and III B.

Finally, we review the results obtained from different measurement techniques in Sec. V. Conclusions follow under Sec. VI.

## II. FROM COATING NOISE TO COATING LOSS ANGLES: THE THERMAL NOISE INTERFEROMETER

As mentioned above, most coating-material characterizations have been done by measuring the mechanical quality factor and then predicting the mechanical noise using the fluctuation-dissipation theorem. Direct interferometric measurements of coating noise are more challenging and hence rarer. The first measurement of this kind was described by Numata [16], referring to a proof-of-principle experiment using intentionally noisy coatings to make the measurement easier. The second direct measurement was done in an apparatus that had been in development longer than Numata's but sought to measure the substantially lower noise floor of the actual coatings used in GW detectors at the time. In addition, it used a larger illumination spot size on the mirrors to further reduce the noise floor. This apparatus was based at Caltech and was known as the Thermal Noise Interferometer (or TNI) (see [17] for details). Conceptually similar instruments are presently under development at the Albert Einstein Institute for Gravitational Physics (Golm and Hannover, Germany) [18] and the University of Florida (Gainesville, Florida, USA) [19], but as of this writing, they have yet to produce useful results. In this paper, we focus on the results from the TNI.

Using the procedure described in [17] and [20], we measured the loss angles of four different coatings at the TNI. From these four independent measurements, we extracted the loss angles of the three relevant coating materials: Silica ($SiO_2$), Tantala ($Ta_2O_5$), and Tantala doped with Titania to a concentration of ~15% [21].

The first coating was a standard quarter wavelength (QWL) stacked-doublets design, using Silica and Tantala for the low and high index materials, respectively.

The second coating also used Silica and Tantala, but the thickness and number of the layers were adjusted so as to minimize thermal noise while keeping the coating reflectivity at the operating wavelength of 1064 nm unchanged. The relevant optimization procedure and the TNI measurements made on this "optimized" coating are described in detail in [20]. The third coating was also QWL and used Silica for the low index material and Tantala doped with Titania to a concentration of ~15% for the high index material [22]. Finally, the fourth coating was designed for minimal noise *dichroic* operation, featuring some extra reflectance at 532 nm needed for locking acquisition in Advanced LIGO, using Silica and the same Titania-doped Tantala. All coatings were deposited on similar fused Silica substrates by ion-beam sputtering. The first coating was manufactured by REO (Research Electro-Optics Inc., Boulder, Colorado, USA), the remaining three by LMA (Laboratoire des Materiaux Avances of the IN2P3, Lyon, France).

TABLE I. The four different coatings whose loss angles were measured at the Caltech LIGO-Lab TNI.

| Coating number | Type | Materials | Manufacturer |
|---|---|---|---|
| 1 | QWL | $SiO_2/Ta_2O_5$ | REO |
| 2 | Optimized | $SiO_2/Ta_2O_5$ | LMA |
| 3 | QWL | $SiO_2/TiO_2::Ta_2O_5$ | LMA |
| 4 | Dichroic optimized. | $SiO_2/TiO_2::Ta_2O_5$ | LMA |

The design type, material composition, and manufacturer of the four coatings are summarized in Table I.

## III. FROM $\phi_C$ TO MATERIAL LOSS ANGLES

Dissipation due to internal friction in a material can be described in terms of a *loss angle* $\phi$, that is, the phase of the material's complex Young's modulus, $\tilde{Y}$. For the materials of interest here, $\phi \ll 1$, and the complex Young's modulus can be written $\tilde{Y} = Y(1 + \iota\phi)$, where $Y$ is the material's elastic (tensile) modulus.

The Power Spectral Density (henceforth PSD) of the coating Brownian noise is related to the effective coating loss angle $\phi_c$ by [23]

$$S_B(f) = \frac{2k_BT}{\pi^{3/2}f}\frac{(1-\sigma_s^2)}{wY_s}\phi_c, \quad (1)$$

where $k_B$ is Boltzmann's constant, $T$ the absolute temperature, $w$ the effective laser Gaussian beam radius, $\sigma_s$ the Poisson's ratio of the substrate, and $Y_s$ its Young's modulus. The effective coating loss angle, $\phi_c$, is a thickness-weighted average of the loss angles of its low and high index constituents [20], viz.

$$\phi_c = b_L d_L \phi_L + b_H d_H \phi_H, \quad (2)$$

where $d_L$ and $d_H$ are the total thickness of the low and high index materials, respectively, $\phi_L$ and $\phi_H$ their loss angles, and the coefficients $b_{L,H}$ are given by

$$b_{L,H} = \frac{1}{\sqrt{\pi}w}\left(\frac{Y_{L,H}}{Y_s} + \frac{Y_s}{Y_{L,H}}\right), \quad (3)$$

$Y_s$, $Y_L$, and $Y_H$ denoting the Young's moduli of the substrate, low index, and high index material, respectively. In the limit of vanishingly small Poisson's ratios [9], Eq. (2) agrees well with the more complicated formula for coating noise derived in [23] from first principles.

Given two coatings, denoted with superscripts (I) and (II), using the same materials but with different thicknesses, Eq. (2) yields

$$\mathbf{M} \cdot \boldsymbol{\phi} = \boldsymbol{\phi}_c, \quad (4)$$

TABLE II. Parameters of the retrieved Gaussian loss angle distributions for the coatings in Table I.

| Coating number | $\mu_c$ | Standard deviation $\sigma_c$ |
|---|---|---|
| 1 | $8.25 \times 10^{-6}$ | $0.3 \times 10^{-6}$ |
| 2 | $6.85 \times 10^{-6}$ | $0.2 \times 10^{-6}$ |
| 3 | $6.0 \times 10^{-6}$ | $0.5 \times 10^{-6}$ |
| 4 | $5.5 \times 10^{-6}$ | $0.25 \times 10^{-6}$ |

where

$$M = \begin{bmatrix} b_L d_L^{(I)} & b_H d_H^{(I)} \\ b_L d_L^{(II)} & b_H d_H^{(II)} \end{bmatrix}, \quad \boldsymbol{\phi} = \begin{bmatrix} \phi_L \\ \phi_H \end{bmatrix}, \quad \boldsymbol{\phi}_c = \begin{bmatrix} \phi_c^{(I)} \\ \phi_c^{(II)} \end{bmatrix}. \quad (5)$$

The low and high index material loss angles are accordingly related to the loss angles of two coatings I and II by an affine (in particular linear) relation,

$$\boldsymbol{\phi} = M^{-1} \cdot \boldsymbol{\phi}_c. \quad (6)$$

In [20] it was noted that the residuals of the fitting used to estimate the coating loss angles from the measured Brownian noise spectra were Gaussian distributed (see Fig. 13 in [20]). The average $\mu_c$ and standard deviation $\sigma_c$ of the estimated loss angle distributions of all coatings in Table I are listed in Table II.

Hence, Eq. (6) yields a jointly Gaussian distribution for $\phi_L$, $\phi_H$ [24]. The related *marginal* distributions of $\phi_L$ and $\phi_H$, which are the quantities of interest, will be Gaussian too and hence completely characterized by their averages $\mu_{L,H}$ and standard deviations $\sigma_{L,H}$, which can be written explicitly (see Appendix).

### A. Silica and Tantala loss angles

The mechanical loss angles of Silica and undoped Tantala were estimated from the noise measurements made on coatings 1 and 2 in Table I. To calculate the elements of $M$, we used the fiducial values $Y_s = Y_L = Y_{\text{SiO}_2} = 73$ GPa and $Y_H = Y_{\text{Ta}_2\text{O}_5} = 140$ GPa for the tensile Young's moduli, used throughout in the topical literature and originated in [25] and the thickness values collected in Table III below.

The total coating thickness uncertainties are of the order of a few $nm$, due to the high accuracy of the coating deposition process, and have no sensible effect on the retrieved material loss angles. On the other hand, as further discussed in Sec. VI, the actual values of the Young moduli may differ from the quoted fiducial ones by a few percent, depending, e.g., on the thermal annealing treatment of the materials. This entails comparable uncertainties in the retrieved material loss angles.

The first- and second-order moments $\mu$ and $\sigma$ of the estimated marginal distributions of $\phi_{\text{SiO}_2}$ and $\phi_{\text{Ta}_2\text{O}_5}$ are collected in Table IV.

TABLE IV. Silica and Tantala loss angles from coatings 1 and 2.

| Loss angle | $\mu$ | $\sigma$ | From error propagation |
|---|---|---|---|
| $\phi_{\text{SiO}_2}$ | $5.14 \times 10^{-5}$ | $2.1 \times 10^{-5}$ | $(5.14 \pm 3.0) \times 10^{-5}$ |
| $\phi_{\text{Ta}_2\text{O}_5}$ | $4.72 \times 10^{-4}$ | $0.43 \times 10^{-4}$ | $(4.72 \pm 0.56) \times 10^{-4}$ |

It is interesting to compare the confidence intervals obtained above, based on the observed Gaussianity of the coating loss angle fitting residuals to the uncertainty intervals obtained from a plain error propagation formula, viz. [26]

$$\delta\boldsymbol{\phi} = \text{abs}(M^{-1}) \cdot \delta\boldsymbol{\phi}_c. \quad (7)$$

The uncertainty intervals obtained from Eq. (7) on letting $\delta\boldsymbol{\phi}_c = \sigma_c$ are also listed in Table IV.

### B. Titania-doped Tantala loss angle

For coatings 3 and 4 in Table I, the matrix $M$ turns out to be ill conditioned, and Eq. (6) yields exceedingly broad confidence intervals.

However, we may safely assume the loss angle of the low-index material (Silica) to be *the same* for *all* coatings in Table I, the low index material being fiducially the same in all. Hence, we may use the Gaussian distribution for $\phi_L$ obtained in Sec. III A to derive from Eq. (2) *two* independent estimates for the loss angle $\phi_{H^*}$ of Titania-doped Tantala from the measured loss angles of coatings 3 and 4. The two distributions can be further *pooled* into a single one (see Appendix for technical details).

The numerical values of the first- and second-order moment of the pooled distribution are collected in Table V.

Similarly, we may obtain *two* uncertainty intervals by applying standard error propagation to Eq. (2) that can be also combined yielding the uncertainty interval in Table V.

TABLE III. Coating structure and total thicknesses of low and high index layers.

| Coating number | Silica layers (nm) | $d_L$ ($\mu$m) | Tantala layers (nm) | $d_H$ ($\mu$m) |
|---|---|---|---|---|
| 1 | $13 \times 181.517 + 1 \times 363.033$ | 2.72 | $14 \times 130.713$ | 1.83 |
| 2 | $16 \times 250.984 + 1 \times 29.410$ | 4.05 | $16 \times 80.688 + 1 \times 72.677$ | 1.36 |
| 3 | $12 \times 181.5 + 1 \times 363.0$ | 2.54 | $13 \times 128.8$ | 1.67 |
| 4 | $12 \times 195.49 + 1 \times 15.48$ | 2.36 | $12 \times 112.10 + 1 \times 103.69$ | 1.45 |

TABLE V. Ti-doped Tantala loss angle from coatings 3 and 4.

| Loss angle | $\mu$ | $\sigma$ | From error propagation |
|---|---|---|---|
| $\phi_{TiO_2::Ta_2O_5}$ | $3.66 \times 10^{-4}$ | $0.27 \times 10^{-4}$ | $(3.6 \pm 0.6) \times 10^{-4}$ |

## IV. COMPARISON WITH AN EMT-BASED MODEL

It is interesting to compare the above results to those obtained from a mixture theory-based approach, also known as effective medium theory (EMT). Despite their simplicity, EMTs admit a solid microscopic foundation [31] and have been widely and successfully used to obtain accurate predictions of the complex refraction index of glassy oxide mixtures, from knowledge (or measurement) of the individual material properties. EMT is valid for inclusions which are small compared to the optical and acoustic wavelengths, and which do not interact to form chemically different compounds. While EMT is admittedly *not* the "Holy Grail" *ab initio* theory that one would like, it emerged as a powerful and versatile tool in Material Science anyway. Its use has been accordingly proposed to model glassy-oxide mixtures for optical coatings [32].

Here we adopt the well-known Bruggemann approach [33], which treats the host medium and the inclusions on equal grounds, assuming both to be embedded into an effective medium, yielding mixture formulas that are symmetric with respect to the host and inclusion parameters. The Bruggemann formula for the (complex) permittivity $\epsilon = n^2$ of a mixture is

$$\eta_2 \frac{\epsilon_2 - \epsilon_{mix}}{\gamma \epsilon_2 + (1-\gamma)\epsilon_{mix}} + (1-\eta_2) \frac{\epsilon_1 - \epsilon_{mix}}{\gamma \epsilon_1 + (1-\gamma)\epsilon_{mix}} = 0, \quad (8)$$

where $\eta$ is the volume fraction, the suffixes 1,2 and mix denote the constituents and the composite, and $\gamma$ depends on the morphology of the inclusions. Here we tentatively adopt the value $\gamma = 3$, appropriate to spherical inclusions. Using the fiducial values $n_{Ta_2O_5} = 2.03$, $n_{TiO_2} = 2.29$, and $n_{TiO_2::Ta_2O_5} = 2.07$, we may use (8) to retrieve the Titania fraction in the doped material, yielding $\eta = 0.16$, as shown in Fig. 1 (top left panel). This value is close to the nominal one for the LMA Ti-doped Tantala [22, 30] used in the coating prototypes tested.

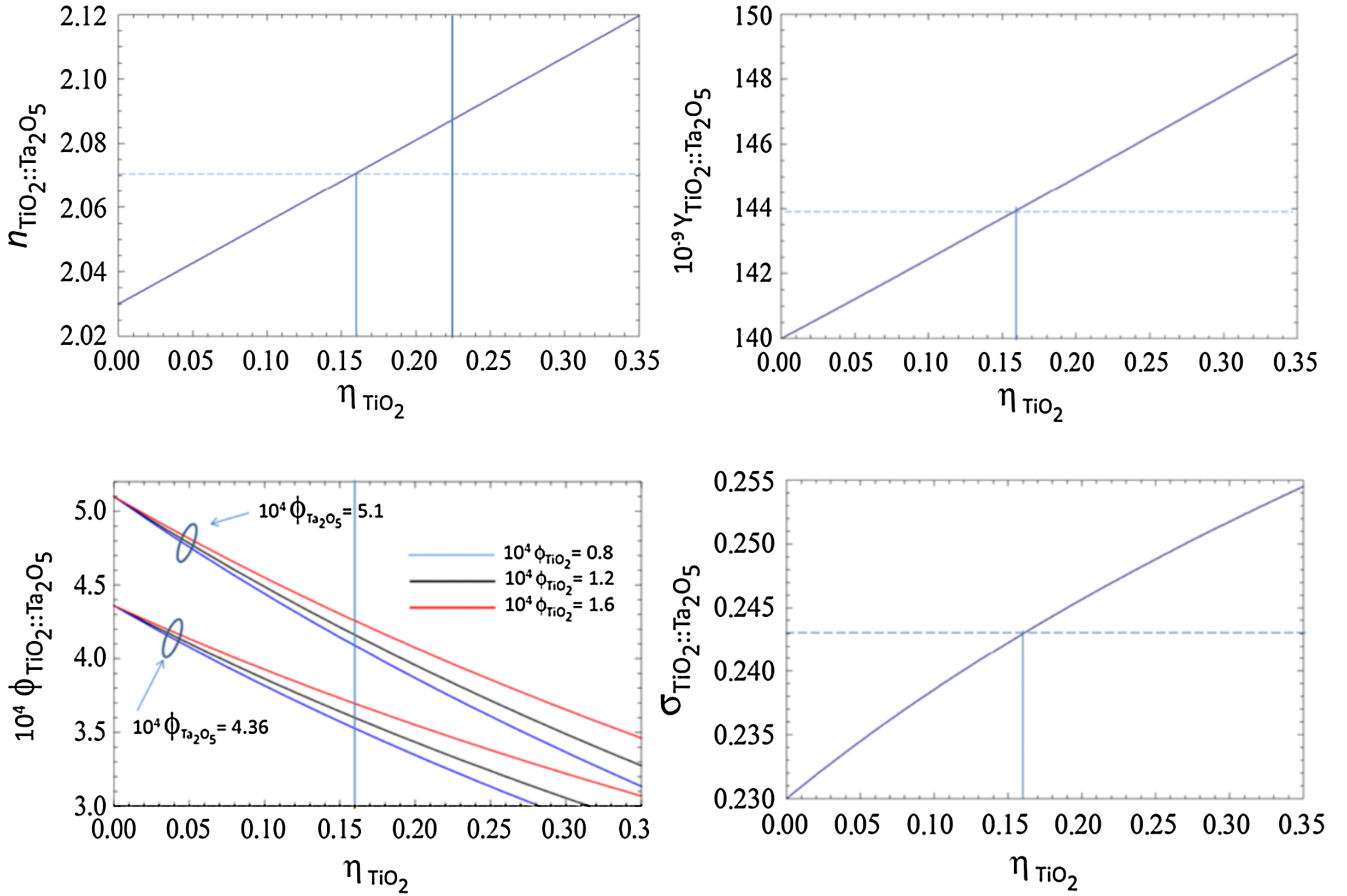

FIG. 1 (color online). Ti-doped Tantala. Refraction index according to Bruggemann formula (top left), tensile (Young) modulus (top right), loss angle (bottom left), and Poisson modulus (bottom right) according to Barta EMT formula.

In order to compute the viscoelastic properties of the mixture, we adopt the physically neat formulation by Barta [34], according to which the complex mixture elastic Young's modulus $Y$ and Poisson's ratio $\sigma$ can be found by solving the system

$$\begin{cases} (1-\eta_2)\frac{X-X_1}{2X+(X_1/y_1)(\sigma_1+1)} + \eta_2 \frac{X-X_2}{2X+(X_2/y_2)(\sigma_2+1)} = 0 \\ (1-\eta_2)\frac{X/y-X_1/y_1}{2X+(X_1/y_1)(\sigma_1+1)} + \eta_2 \frac{X/y-X_2/y_2}{2X+(X_2/y_2)(\sigma_2+1)} = 0 \end{cases}, \quad (9)$$

where omitting the subscripts for notational ease,

$$X = \frac{\sigma Y}{\sigma+1}, \qquad y = \sigma - 2. \quad (10)$$

Equations (9) and (10) can be used to compute the Young's modulus and Poisson's ratio of doped Tantala, using the fiducial values $Y_{Ta_2O_5} = 140$ GPa, $Y_{TiO_2} = 165$ GPa, $\sigma_{Ta_2O_5} = 0.23$, and $\sigma_{TiO_2} = 0.28$. The results are shown in Fig. 1.

The real part of the mixture's Young's modulus and Poisson's ratio (top right and bottom right in Fig. 1) shows no sensible dependence on the very small constituents' loss angles. The loss angle (imaginary part of the elastic modulus) depends on the loss angle of amorphous Tantala and Titania, as shown in the bottom left of Fig. 1.

We next attempt to compute a confidence interval for the Titania-doped Tantala loss angle, computed via EMT Eqs. (9) and (10), assuming for the Tantala loss angle a Gaussian distribution obtained from the TNI measurements on undoped coatings and for the Titania loss angle a Gaussian distribution, with average value $1.2 \times 10^{-4}$ taken from [35] and a reasonable value for the standard deviation of 10% its average value.

The EMT deduced Ti-doped Tantala loss angle distribution is shown in Fig. 2, where it is compared to the (pooled) distribution obtained from our measurements on coatings 3 and 4. The two distributions look fairly consistent. Thus, we have a simple theory that, at least in the present case, predicts fairly well the loss angle of the doped material from the known properties of its components, yielding results that are consistent with experimental observations, within the uncertainties of the measurements.

## V. OTHER MEASUREMENT METHODS AND RESULTS

During the last decade, the mechanical loss angles of various candidate coating materials for interferometric GW detectors have been estimated by several research groups, both at room and cryogenic temperatures, from the measured damping times of mechanical oscillators consisting of thin/thick disk or cantilever shaped blades, before and after coating deposition. This section presents a brief review of available room temperature results, summarized in Table VI, mostly referring to ion-beam sputtered coatings, for comparison.

### A. Suspended Disk Blades

A measurement setup based on suspended disk-shaped thin or thick blades was described in [36,37] and used to estimate the mechanical losses of several glassy oxides. Knowledge of the mechanical and optical losses of candidate materials led to downselect Silica and Tantala as the "best" low and high index materials available for interferometric gravitational wave detector mirror coatings

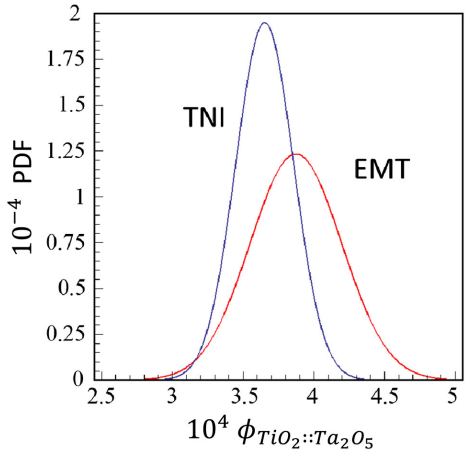

FIG. 2 (color online). Comparison between Titania-doped Tantala loss angle distributions resulting from TNI measurements and EMT.

TABLE VI. Loss angles of different materials from various measurement methods.

| Material | Layer thickness (nm) | $\phi$ ($\times 10^{-4}$) | Source |
| --- | --- | --- | --- |
| SiO$_2$ | 90.8–272.3 | $0.5 \pm 0.3$ | Suspended disks [12] |
| | 181.5–250.984 | $0.51 \pm 0.07$ | TNI |
| | 500 | $0.5 \pm 0.018$ | Clamped cantilevers [27] |
| | 500 | $0.46 \pm 0.01$ | Clamped cantilevers [28] |
| | 3,070 | $0.6 \pm 0.03$ | Quad. Phase Diff. IFO [29] |
| Ta$_2$O$_5$ | 65.36–196.07 | $4.4 \pm 0.2$ | Suspended disks [12] |
| | 80.688–130.713 | $4.72 \pm 0.14$ | TNI |
| | 133 | $3.3 \pm 0.9$ | GeNS [30] |
| | 500 | $3.02 \pm 0.11$ | Clamped cantilevers [27] |
| | 3,130 | $4.7 \pm 0.2$ | Quad. Phase Diff. IFO [29] |
| TiO$_2$:Ta$_2$O$_5$ | 112.10–128.8 | $3.66 \pm 0.26$ | TNI |
| | 500 | $2.4 \pm 0.4$ | Clamped cantilevers [28] |

[38]. The main results obtained using this setup were summarized in [12]. One of the main results was that noise originated mainly from the coating bulk, the interfacial contributions being negligible. Also, the following estimates for the loss angles of annealed $SiO_2$ and un-doped $Ta_2O_5$ were given $\phi_L = (0.5 \pm 0.3) \times 10^{-4}$ and $\phi_H = (4.4 \pm 0.2) \times 10^{-4}$ at frequencies $\sim 10^3$ Hz.

These values, as reported in [12], are consistent with ours as reported in Table IV for both Silica and undoped Tantala. However, the authors reanalyzed their data in a later publication [38], and their amended values are not consistent with our results. It is worth noting that the thicknesses of the samples measured in [12,36–38] vary from $\lambda/8$ to $3\lambda/8$ and are thus in the same general range as our layer thicknesses, which were $\lambda/4$ for coating 1 and $0.62\lambda/4$ for coating 2.

### B. Clamped Cantilevers

A different setup, based on clamped cantilever-shaped blades, was developed at LMA, in collaboration with researchers from the Universities of Perugia and Glasgow. An analytic model of the cantilever oscillator allowing to extract the coating loss angles from the measured quality factors was laid out in [39] for single-layer coatings and in [27] for the multilayer ones.

This setup was used to estimate the loss angle of cantilevers coated with a single layer of Silica or (undoped) Tantala, at frequencies $\sim 10^2$ Hz, yielding $\phi_L = (0.5 \pm 0.018) \times 10^{-4}$ and $\phi_H = (3.02 \pm 0.11) \times 10^{-4}$, respectively [27].

The same setup was used at LMA to optimize *mixtures* where Tantala was doped with different materials, including Cobalt, Tungsten, and Titanium, to reduce its mechanical losses [27]. It was found that $Ta_2O_5$ doped with Ti at concentrations $\approx 14\%$ was almost as good as undoped Tantala in terms of optical absorption, but better by $\approx 17\%$ in terms of loss angle. A consistent reduction in loss angle going from plain to Ti-doped Tantala was observed also using a suspended disk Q-measurement setup [22] and also from TNI measurements. Experiments on other doped oxides (in particular $ZrO_2$) at LMA eventually indicated that Ti-doped Tantala was the best option for the high index material [40], among the tested ones.

Mesurements on single-layer coated cantilevers from several groups produced consistent results for the Silica and Ti-doped Tantala loss angles [28], yielding $\phi_L = (4.6 \pm 0.1) \times 10^{-5}$ and $\phi_{H^*} = (2.4 \pm 0.4) \times 10^{-4}$, denoting here and henceforth the loss angle of Ti-doped Tantala as $\phi_{H^*}$.

Results for $SiO_2$ are consistently in agreement with our results, but the results for both doped and undoped Tantala are *not* consistent with our results as listed in Tables IV and V, based on direct noise measurements. It is worth noting that the thicknesses of the individual Tantala layers in these clamped-cantilever measurements are 500 nm compared with 132 nm in our coatings.

### C. Multilayer coated cantilevers

Loss angle measurements on multilayer-coated cantilevers started around the year 2009. Coating loss angles *larger* than those extrapolated from single-layer results were obtained. The origin of the observed excess noise is, as yet, unclear.

Assuming $\phi_{H^*} = (2.4 \pm 0.2) \times 10^{-4}$, the multilayer cantilever based measurements yield $\phi_L = (1.3 \pm 0.4) \times 10^{-4}$, significantly larger than the value ($\approx 5 \times 10^{-5}$) retrieved from single-layer Silica-coated blades [41]. On the other hand, assuming $\phi_L \approx 0.5 \times 10^{-4}$, the same multilayer cantilever-based measurements yield $\phi_H = (4.2 \pm 0.2) \times 10^{-4}$, much larger than the value ($\approx 2.4 \times 10^{-4}$) retrieved from single-layer Titania-doped-Tantala-coated blades [41], and consistent with our results.

Further measurements at LMA indicated that excess noise was increasing with the number of layers [41], suggesting that excess losses could originate at the interfaces between the high and low index layers, in disagreement with results in [12] based on suspended multilayer coated disk measurements.

It was also suggested that interfacial diffusion during the annealing phase, producing graded/index regions at the boundaries between the low and high index layers, may account for the observed discrepancy [42]. A subsequent analysis based on EMT shows that interfacial diffusion is not sufficient to contribute the observed extra noise [43].

It was further observed that the distribution of the loss-angle fitting residuals of cantilever-based loss angle measurements is usually markedly non-Gaussian [44]. Robust estimation of the retrieved loss-angle confidence intervals would be accordingly in order, possibly mitigating the noted discrepancies between loss angle estimates based on single-layer and multilayer blades.

### D. The Gentle Nodal Suspension

The accuracy and repeatability of clamped-cantilever-based measurement are severely affected by clamping losses. Reducing these latter requires careful control of the contacting surfaces of the clamping vise and cantilever [28]. These problems can be effectively mitigated using a different setup, where a disk-shaped blade is supported at a nodal point of its mechanical vibration pattern by a hard (e.g., sapphire) conical tip, ideally without friction [45].

Ringdown measurements of single-layer undoped Tantala-coated silicon disks, based on this setup, nicknamed *GeNS*, for *Ge*ntle *N*odal *S*uspension, yield loss angle values $\phi_H = (3.3 \pm 0.9) \times 10^{-4}$ for 133 nm monolayers of $Ta_2O_5$, with very good repeatability [30,45]. This result is consistent, to within two standards deviations, with our results in Table IV. Measurements on Ti-doped Tantala are underway.

### E. Quadrature phase differential interferometry

A different measurement setup for the direct measurement of broadband thermal noise of coated cantilevers based on quadrature phase differential interferometry [46] has been described in [29,47].

The loss angles of $SiO_2$ and undoped $Ta_2O_5$ estimated from these measurements were $\phi_L = (6.0 \pm 0.1) \times 10^{-5}$ and $\phi_H = (4.7 \pm 0.2) \times 10^{-4}$. Both values are reasonably consistent with our results as reported in Table IV, even though the thicknesses of their samples were much larger than ours $(3.07 \pm 0.12)$ $\mu$m of Silica and $(3.13 \pm 0.12)$ $\mu$m of Tantala versus 182 and 131 nm, respectively. Measurements on Ti-doped Tantala are underway.

### F. Young's modulus

Retrieving the material loss angles from the measured loss angles of disks/blades before and after coating relies on knowledge of the ratio, known as the energy dilution factor, between the energies stored in the coating and substrate [12,27,39,47]. This latter, can be expressed in terms of the tensile (Young) moduli of the substrate and coating materials [48].

The fiducial estimates $Y_L = 72.7$ GPa and $Y_H = 140$ GPa for Silica and (Titania doped as well as undoped) Tantala, respectively, taken from optical glass databases, have been widely used for this purpose. Accurate values of the Young moduli are also needed to retrieve the material loss angles from the coating ones, as in Sec. III of the present paper.

Accurate measurements of the tensile Young's modulus based both on nano indentation [49] and ultrasonic reflection techniques [50] are ongoing. Preliminary results in [51] indicate that the Young's modulus for Titania-doped Tantala may vary in a rather wide range, roughly from 120 to 175 Gpa, depending on dopant concentration and heat treatment.

## VI. CONCLUSIONS

Accurate measurements of the viscoelastic properties of glassy oxides are needed to design better coatings for GW detectors. This a relatively recent research field, no older than 12 years. Experimental setups for material loss angle and Young's modulus measurements have been steadily improving, resulting into better and better accuracy and repeatability.

Sofar, material losses have been estimated from mechanical Q measurements. In this paper, we presented a derivation of the individual material loss angles, including pertinent uncertainties, from the direct measurement of thermal noise in the mirror coatings of an interferometer, in a frequency range relevant to interferometric gravitational-wave detectors.

During the review process, we became aware of a recent work by Chalermsongsak *et al.*, where direct noise measurements from a new rigid cavity instrument are combined with early ringdown measurements in a Bayesian perspective [52], similar to ours.

We also presented here a simple, *predictive* theory for the material properties of glassy oxide mixtures, based on EMT. All approaches to mixture optimization proposed so far required fabrication first, followed by measurement of the relevant optical and mechanical properties. Our simple approach reproduces accurately our measured values of the loss-angle of Ti-doped Tantala.

As of today, loss angle estimates from different measurement methods and facilities exhibit non-negligible discrepancies. The reasons of such discrepancies are yet unclear. A number of possible causes have been scrutinized so far, without conclusive results. Ongoing efforts toward better knowledge of the relevant process/dependent material parameters (in particular, the Young's modulus) and improved coating-noise models may hopefully help in clarifying these issues.

We believe that the present work adds to the available body of knowledge and will stimulate further investigations.

## ACKNOWLEDGMENTS

This work has been supported in part by the NSF under Cooperative Agreement No. PHY-0757058 and the Italian National Institute for Nuclear Physics (INFN) under the CSN-V COAT grant. Stimulating discussions with V. B. Braginsky, G. Cagnoli, R. Flaminio, P. Fritschel, G. Harry, K. Kuroda, J. M. Mackowski, I. Martin, N. Mio, and S. Penn are gratefully acknowledged.

## APPENDIX: MATERIAL LOSS ANGLES' DISTRIBUTIONS

In view of Eq. (4), if we model $\phi_c^{(I)}$ and $\phi_c^{(II)}$ as independent Gaussian random variables with known averages $\mu_c^{(I,II)}$ and standard deviations $\sigma_c^{(I,II)}$, then $\phi_L$ and $\phi_H$ will be jointly Gaussian [24], and their distribution $\Psi_2(\phi_L, \phi_H)$ will be completely characterized by the average vector

$$\boldsymbol{M}^{-1} \cdot E[\boldsymbol{\phi}_c] \quad (A1)$$

and the covariance matrix

$$\boldsymbol{M}^{-1} \cdot \begin{bmatrix} \sigma^2_{\phi_c^{(I)}} & 0 \\ 0 & \sigma^2_{\phi_c^{(II)}} \end{bmatrix} \cdot [\boldsymbol{M}^{-1}]^T. \quad (A2)$$

The joint distribution of $\phi_{SiO_2}$ and $\phi_{Ta_2O_5}$, obtained from Eqs. (A1) and (A2) using the measured loss angles of coatings 1 and 2, is shown in Fig. 3 (left panel), together with a few of its quantile ellipses (right panel). These latter are *squeezed* along a line going through the point $\{E(\phi_L), E(\phi_H)\}$ where the distribution is peaked, with

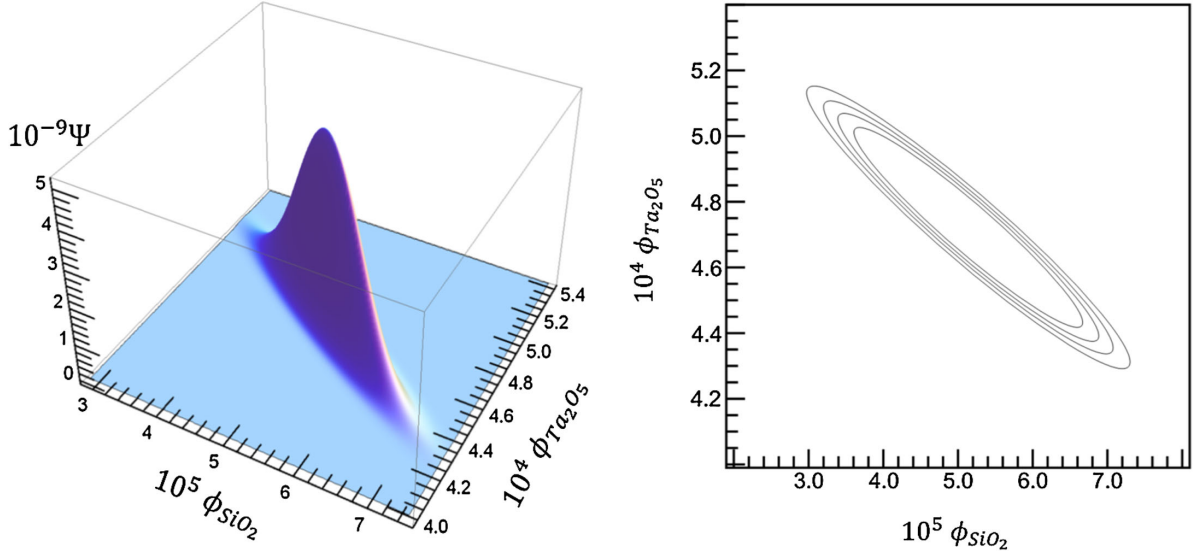

FIG. 3 (color online). (Left) Joint distribution of $\phi_{SiO_2}$ and $\phi_{Ta_2O_5}$ obtained from the measured loss angles of coatings 1 and 2 in Table I. (Right) The 0.95, 0.9, 0.85 quantile ellipses of the same distribution.

slope $\approx -0.51$ reflecting the correlation between $\phi_L$ and $\phi_H$, represented by the nondiagonal matrix (A2). The *marginal* distributions of $\phi_L$ and $\phi_H$,

$$\Psi_L(\phi_L) = \int_{-\infty}^{\infty} d\phi_H \Psi_2(\phi_L, \phi_H),$$

$$\Psi_H(\phi_H) = \int_{-\infty}^{\infty} d\phi_L \Psi_2(\phi_L, \phi_H), \quad (A3)$$

are readily computed in closed analytic form and, being Gaussian are completely characterized by their means and standard deviations, used to obtain the numbers in the middle column of Table IV and given by

$$\mu_{L,H} = \frac{\mu_c^{(I)} d_{H,L}^{(II)} - \mu_c^{(II)} d_{H,L}^{(I)}}{b_{L,H}(d_{L,H}^{(I)} d_{H,L}^{(II)} - d_{L,H}^{(II)} d_{H,L}^{(I)})} \quad (A4)$$

$$\sigma_{L,H}^2 = \frac{(d_{H,L}^{(II)} \sigma_c^{(I)})^2 + (d_{H,L}^{(I)} \sigma_c^{(II)})^2}{b_{L,H}^2 (d_{L,H}^{(I)} d_{H,L}^{(II)} - d_{L,H}^{(II)} d_{H,L}^{(I)})^2}. \quad (A5)$$

Standard error propagation, Eq. (7), is equivalent to the simple graphic construction shown in Fig. 4, where the uncertainty intervals follow from the intersections of the uncertainty *strips* in the $\{\phi_H, \phi_L\}$ plane obtained from Eq. (2) upon letting $\phi_c = \mu_c^{(I,II)} \pm \sigma_c^{(I,II)}$.

For coatings 3 and 4 in Table I, the matrix **M** turns out to be ill conditioned, and Eqs. (A1)–(A3) yield exceedingly broad confidence intervals.

The low-index material (Silica) being fiducially the same for *all* coatings, we may use the Gaussian distribution for $\phi_L$ obtained from coatings 1 and 2, in Eq. (2) to derive *two* (independent) estimates for the loss angle $\phi_{H^*}$ of the Titania-doped Tantala, from the measured loss angles of coatings 3 and 4. The high-index material loss angles retrieved from Eq. (2) will be Gaussian distributed, with

$$E[\phi_{H^*}] = \frac{1}{b_H d_H} \mu_c - \frac{b_L d_L}{b_H d_H} \mu_L, \quad (A6)$$

$$\text{var}[\phi_{H^*}] = \left(\frac{1}{b_H d_H}\right)^2 \sigma_c^2 + \left(\frac{b_L d_L}{b_H d_H}\right)^2 \sigma_L^2. \quad (A7)$$

The *two* distributions obtained from coatings 3 and 4, henceforth labeled with the suffixes 3 and 4, can be further combined (technically, pooled, or conflated [53]) to obtain

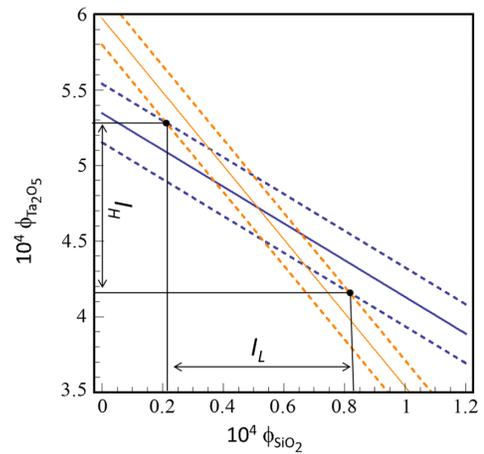

FIG. 4 (color online). Graphic construction for standard error propagation for coatings 1 and 2, showing the intersection between the uncertainty strips obtained from Eq. (2) upon letting $\phi_c = \mu_c^{(I,II)} \pm \sigma_c^{(I,II)}$. The resulting uncertainty intervals for SiO$_2$, $I_L$, and Ta$_2$O$_5$, $I_H$ are indicated.

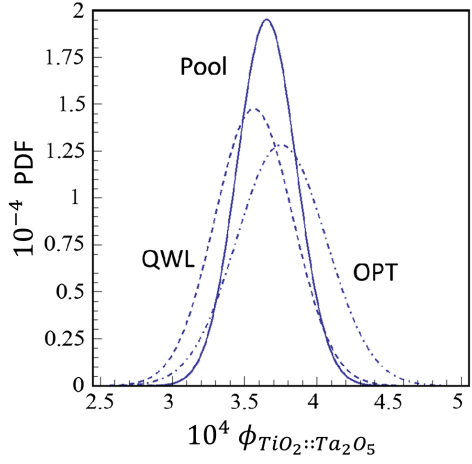 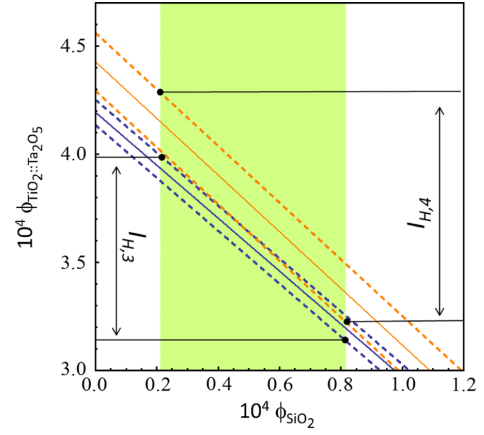

FIG. 5 (color online). Titania doped Tantala loss angle distributions from coatings 3 (QWL) and 4 (OPT) and their pooled (maximum likelyhood) combination. The average and standard deviation of the pooled distribution are $3.66 \times 10^{-5}$ and $0.26 \times 10^{-5}$, respectively.

FIG. 6 (color online). Graphic construction for standard error propagation for coatings 3 and 4. The red and blue strips are obtained from Eq. (2) using the measured loss angles of coatings 3 and 4 and their uncertainties. The green band is the Silica loss angle uncertainty strip obtained from measurements on coatings 1 and 2. The intersection of the green band with each coating measurement yields two uncertainty intervals for $TiO_2::Ta_2O_5$ loss angle, $I_{H,3}$ and $I_{H,4}$, respectively. The pooled uncertainty interval is $I_{H,3} \cap I_{H,4}$.

a (Gaussian) maximum-likelihood distribution for $\phi_{H^*}$ whose first- and second-order moments are [24]

$$E[\phi_{H^*}] = w_3 E[\phi_{H^*}]_3 + w_4 E[\phi_{H^*}]_4, \quad (A8)$$

$$\mathrm{var}[\phi_{H^*}] = \frac{1}{2}(w_3 \mathrm{var}[\phi_{H^*}]_3 + w_4 \mathrm{var}[\phi_{H^*}]_4) \quad (A9)$$

where

$$w_{3,4} = \frac{\mathrm{var}[\phi_{H^*}]_{3,4}^{-1}}{\mathrm{var}[\phi_{H^*}]_3^{-1} + \mathrm{var}[\phi_{H^*}]_4^{-1}}. \quad (A10)$$

Note that Eq. (A8) is also the best linear unbiased estimator of $\phi_{H^*}$. The two distributions obtained from coatings 3 and 4 and their pooled combination are shown in Fig. 5. Standard error propagation is equivalent in this case to first intersecting *each* of the loss uncertainty strips, obtained through Eq. (2), for coatings 3 and 4, with the strip $\phi_L = \mu_L \pm \sigma_L$, and *then* computing the intersection of the resulting uncertainty intervals for $\phi_{H^*}$, as shown in Fig. 6. This corresponds to assuming, in the spirit of plain error propagation, a uniform distributon of $\phi_{H^*}$ in the two uncertainty intervals in Fig. 6 and constructing the conflated (pooled) distribution [53].